\documentclass[onecolumn,secnumarabic,amsmath,amssymb,balancelastpage,nofootinbib]{article}

\usepackage[e]{esvect}
\usepackage{color}         
\usepackage{graphics}      
\usepackage{graphicx}      
\usepackage{epsf}          
\usepackage{bm}            

\usepackage{natbib}
\usepackage{amssymb}
\usepackage{amsmath, esint}
\usepackage{mathrsfs}
\usepackage{framed}
\usepackage{bigints} 
\usepackage{enumitem}
\usepackage{pifont}
\usepackage[capbesideposition={left,center},facing=yes,capbesidewidth=8cm,capbesidesep=quad]{floatrow} 
\usepackage{setspace} 

\usepackage[none]{hyphenat} 

\usepackage[colorlinks=true]{hyperref}  

\definecolor{darkred}{rgb}{0.6,0,0}
\definecolor{darkgreen}{rgb}{0,0.5,0}
\definecolor{darkblue}{rgb}{0,0,0.6}
\hypersetup{ colorlinks,
linkcolor=darkblue,
filecolor=darkgreen,
urlcolor=darkgreen,
citecolor=darkred }

\newcommand{\del}[0]{\ensuremath{\vec{\nabla}}}

\begin{document}

\sloppy 

\bibliographystyle{authordate1}


\title{\vspace*{-35 pt}\Huge{Electromagnetism as\\Quantum Physics}}
\author{Charles T. Sebens\\California Institute of Technology}
\date{May 29, 2019\\ arXiv v.3\\\vspace*{6 pt} {\normalsize The published version of this paper appears in} \\ \vspace*{-3 pt}{\normalsize \textit{Foundations of Physics}, 49(4) (2019), 365-389.} \\{\small\url{https://doi.org/10.1007/s10701-019-00253-3}}}


\maketitle
\vspace*{-20 pt}
\begin{abstract}
One can interpret the Dirac equation either as giving the dynamics for a classical field or a quantum wave function.  Here I examine whether Maxwell's equations, which are standardly interpreted as giving the dynamics for the classical electromagnetic field, can alternatively be interpreted as giving the dynamics for the photon's quantum wave function.  I explain why this quantum interpretation would only be viable if the electromagnetic field were sufficiently weak, then motivate a particular approach to introducing a wave function for the photon (following Good, 1957).  This wave function ultimately turns out to be unsatisfactory because the probabilities derived from it do not always transform properly under Lorentz transformations.  The fact that such a quantum interpretation of Maxwell's equations is unsatisfactory suggests that the electromagnetic field is more fundamental than the photon.
\end{abstract}

\tableofcontents
\newpage

\section{Introduction}

Electromagnetism was a theory ahead of its time.  It held within it the seeds of special relativity.  Einstein discovered the special theory of relativity by studying the laws of electromagnetism, laws which were already relativistic.\footnote{More accurately: Maxwell's equations and the Lorentz force law giving the force exerted upon a charged body were already special relativistic, though the reaction of bodies to forces from the electromagnetic field must be understood properly (as the rate of change of \emph{relativistic} momentum) if the theory as a whole is to be relativistic.}  There are hints that electromagnetism may also have held within it the seeds of quantum mechanics, though quantum mechanics was not discovered by cultivating those seeds.  Maxwell's equations are generally interpreted as classical laws governing the electromagnetic field.  But, if the electromagnetic field is sufficiently weak, it may be possible to interpret the same equations as quantum laws governing the photon's wave function.

Finding such an interpretation would bring the physics of the photon into closer alignment with the physics of the electron.  Consider Dirac's relativistic wave equation for the electron,\footnote{See \citet[sec.\ 1.3]{bjorkendrell}.}
\begin{equation}
i\hbar \frac{\partial \psi}{\partial t}=\left(c\:\vec{\alpha}\cdot\vec{p}+\beta m c^2 \right)\psi
\ .
\label{thediracequation}
\end{equation}
We can either interpret $\psi$ as a classical field and the Dirac equation as part of a relativistic classical field theory, or, we can interpret $\psi$ as a single-particle wave function and the Dirac equation as part of a relativistic quantum theory of the electron.\footnote{The Dirac equation is treated as part of a classical field theory in, e.g., \citet{hatfield}; \citet[ch.\ 4]{valentini1992}; \citet{peskinschroeder}; \citet[sec.\ 4.3]{ryder1996}; \citet[ch.\ 5]{greiner1996}; \citet{howelectronsspin} and it is treated as part of a quantum particle theory in, e.g., \citet{frenkel, dirac, schweberQFT, bjorkendrell, thaller1992}.  There is much to be said about how one moves from either one of these options to a full quantum field theory.  However, as the focus of this paper is on the physics of the photon, I will not say much about that here.  Let me just mention that at some point along the road to quantum field theory, one must shift to thinking of Dirac's equation as part of a theory of both electrons and positrons.}  Neither of these theories is a quantum field theory, but each comes with its own path to quantum field theory.  The first interpretation of \eqref{thediracequation} fits with a field approach to quantum field theory where the next step would be to quantize the classical Dirac field.\footnote{Classical Dirac field theory has not proved useful as a theory of macroscopic physics.  This is because, unlike classical electromagnetism, classical Dirac field theory does not emerge as a classical limit of quantum field theory \citep[pg.\ 221]{duncan}.  However, the theory is still of interest to physics because, like classical electromagnetism, classical Dirac field theory serves an important foundational role in a field approach to quantum field theory (as mentioned above).}  The second interpretation of \eqref{thediracequation} fits with a particle approach to quantum field theory where one would next move from this single-particle quantum theory to quantum field theory by extending the theory to multiple particles and allowing for superpositions of different numbers of particles.

The availability of both field and particle interpretations of the Dirac equation makes it difficult to determine whether we should take a field or particle approach to our quantum field theory of the electron (whether we should take the Dirac field or the electron to be more fundamental).\footnote{The question of whether we should take a field or a particle approach to various quantum field theories has been a popular topic of discussion in the philosophy of physics literature.  See, e.g., \citet{malament1996, halvorsonclifton2002, fraser2008, baker2009, baker2016, struyve2011, wallace2017}.}  If, like the Dirac equation, Maxwell's equations can be interpreted either as part of a relativistic classical field theory or a relativistic quantum particle theory, then we face a similar puzzle about whether we should take the electromagnetic field or the photon to be more fundamental.  If, on the other hand, we cannot find a satisfactory interpretation of Maxwell's equations as part of a relativistic quantum particle theory, then that points toward the electromagnetic field being more fundamental than the photon.

There are a number of ways to make electromagnetism look like quantum physics and, correspondingly, a number of different mathematical objects built from $\vec{E}$ and $\vec{B}$ that might deserve to be called the wave function of the photon.  We will see that one common proposal, the Weber vector, is quite simple and already looks very much like a quantum wave function.  However, its square gives an energy density (or perhaps an energy-weighted probability density), not a straight probability density.  What I will call the photon wave function is a less common and slightly more complicated construction from the electromagnetic field, originally proposed by \citet{good1957}.  Given that one rarely sees any attempt to find a wave function for the photon in physics textbooks, I think it is remarkable that Good's proposal works as well as it does.  However, this wave function is ultimately unsatisfactory as a \emph{relativistic} wave function for the photon because the probabilities generated from it do not transform properly under Lorentz transformations---as will be shown explicitly in section \ref{EXsection}.  The formulas for probability density and probability current are not part of the classical field theory of electromagnetism and break that theory's fit with special relativity.  Although it may be possible to further manipulate Maxwell's equations to find a fully relativistic wave function for the photon, I am not optimistic.  The photon wave function examined here already looks very much like the Dirac wave function for the electron.  Thus, I take the lesson of this analysis to be that: although it is possible to make electromagnetism look very much like a quantum theory, it is better interpreted as a field theory.

In the next section I use the Weber vector to present electromagnetism in a way that closely resembles quantum mechanics.  Next, I apply the Planck-Einstein relation between the energy and frequency of a photon to determine how weak the electromagnetic field would have to be for it to describe the state of a single photon.  I then build on insights from that analysis to introduce Good's photon wave function as a natural wave function for the photon.  After giving the motivation behind Good's photon wave function, I expose its primary flaw by going through a number of examples to show that: although the probabilities derived from the photon wave function sometimes transform properly under Lorentz transformations, they don't always do so.  Thus, observers moving relative to one another might have radically different expectations as to the results of position measurements even when they agree on the initial state.  This is unacceptable for a relativistic quantum theory.

While we are looking within the equations of electromagnetism for a quantum theory giving the dynamics of the photon wave function, we will also consider the prospects for developing a Bohmian version of such a theory: positing the existence of an actual photon particle that is separate from the wave function and guided by it.  This is the way that the electron is generally treated by the proponents of Bohmian approaches to quantum physics.  They take the electron's wave function to evolve by the Dirac equation (never undergoing collapse) and the motion of the electron itself to be guided by its wave function in accord with a new law of nature, the guidance equation.\footnote{The relativistic Bohmian dynamics for the electron are presented in \citet{bohm1953}; \citet[sec.\ 12.2]{bohmhiley}; \citet[sec.\ 12.2]{holland}.} The photon, however, is treated quite differently as this sort of road has been thought to be closed off.  Instead of looking for the dynamics of a quantum particle, one starts with a quantum field theory of the electromagnetic field and postulates that in addition to the quantum state (which may be represented as a wave functional) there is a single actual configuration of the electromagnetic field which evolves in accord with a guidance equation linking quantum state and field.\footnote{This field approach dates back to \citet{bohm1952pt2}.  For some discussion as to why the field approach has been taken instead of the particle approach, see \citet[ch.\ 11]{bohmhiley}; \citet{holland1993}; \citet[sec.\ 12.6]{holland}; \citet{struyve2011}; \citet{flack2016}.}  Any particle-like behavior that must be accounted for is to be found in the dynamics of this field.  Although I have nothing to say against such a field approach, I would like to see whether it is indeed the only road available by evaluating potential Bohmian dynamics for the photon.  I will propose a new guidance equation for the photon based on Good's photon wave function.  However, we will see that this guidance equation is ultimately inadequate due to the problems that Good's wave function has with relativity (mentioned above).

Einstein was the one to grow the special theory of relativity from the seeds present in electromagnetism.  He also attempted to grow a quite Bohmian-looking quantum theory of the photon:
\begin{quote}
``[Einstein] was very early well aware of the wave-particle duality of the behavior of light ... In order to explain this duality of their behavior, Einstein proposed the idea of a `guiding field' (\textit{F\"{u}hrungsfeld}).  This field obeys the field equations for light, that is Maxwell's equation.  However, the field serves only to \emph{guide} the light quanta or particles, they move into regions where the intensity of the field is high.  This picture has a great similarity with the present picture of quantum mechanics and has, obviously, many attractive features.  Yet Einstein, though in a way he was fond of it, never published it.'' \citep[pg.\ 463]{wigner1980}\footnote{Einstein's hope for what looks like a Bohmian theory of the photon has been mentioned by \citet[pg.\ 538]{holland}; \citet{kiessling2017}.  See also \citet[pg.\ 440--443]{pais1982}; \citet[pg.\ 262]{wigner1983}.}
\end{quote}
We will search for such a theory, but---like Einstein---we will not be satisfied with the results of that search.

\section{The Weber Vector}\label{RSsection}

It is possible to reformulate classical electromagnetism in a way that resembles quantum mechanics by expressing the state of the electromagnetic field using a single complex vector field\footnote{\citet[ch.\ 1]{lindell1992} discusses the mathematical properties of complex vectors.} $\vec{F}$ (the Weber vector\footnote{\citet[appendix A]{kiessling2017} present a historical case as to why $\vec{F}$ should be called the ``Weber vector'' (after Heinrich Martin Weber) and not the ``Riemann-Silberstein vector'' (a name introduced by Bialynicki-Birula).}),
\begin{equation}
\vec{F}=\vec{E}+i\vec{B}
\ ,
\label{rsvector}
\end{equation}
in place of the separate electric and magnetic fields $\vec{E}$ and $\vec{B}$.  Like $\vec{E}$ and $\vec{B}$, $\vec{F}$ is a function of space and time.  The energy density of the electromagnetic field can be written in terms of the Weber vector as
\begin{equation}
\rho^{\mathcal{E}}=\frac{1}{8 \pi}\vec{F}^{*}\!\cdot\vec{F}=\frac{1}{8 \pi}\left(E^2+B^2\right)
\ ,
\label{energydensityfield}
\end{equation}
where the $\mathcal{E}$ superscript indicates that this is a density of energy.  The energy flux density (Poynting vector) is
\begin{equation}
\vec{S}=\frac{c}{8 \pi i}\vec{F}^*\times\vec{F}= \frac{c}{4\pi} \vec{E} \times \vec{B}
\ .
\label{poyntingvector}
\end{equation}
Here and throughout, Gaussian units are used.

Maxwell's equations in vacuum can be written in terms of $\vec{F}$ as
\begin{align}
i\frac{\partial \vec{F}}{\partial t}  = c\del\times\vec{F}
\label{newmaxwell1}
\\
\del\cdot\vec{F}=0
\ .
\label{newmaxwell2}
\end{align}
Here the similarity to quantum mechanics becomes clear.  The equation of time evolution \eqref{newmaxwell1} looks like a Schr\"{o}dinger equation of the general form
\begin{align}
i \hbar \frac{\partial \psi}{\partial t} = \widehat{H} \psi
\ ,
\end{align}
in which $\vec{F}$ acts like a wave function with its evolution determined by a Hamiltonian $\widehat{H}=\hbar c \del\times$.\footnote{The idea that $\vec{F}$ might be interpreted as a wave function for the photon is explored in: \citet{rumer1930, archibald1955}; \citet{good1957, good1985}; \citet[ch.\ 11]{goodnelson1971}; \citet{moses1959}; \citet{kobe1999, esposito1999, holland2005, raymer2005, keller2005, cugnon2011, chandrasekar2012}; \citet[pg.\ 117]{norsen2017}; \citet{kiessling2017}; and a number of publications by Bialynicki-Birula and Bialynicka-Birula, including \citet{bb1994, bb1996, bb2013}.  The earliest sources for this quantum interpretation of $\vec{F}$ appear to be \citet{rumer1930} and Majorana (who worked on this between 1928 and 1932; see \citealp{mignani1974}).  For more on the role of the Weber vector in classical electromagnetism, see \citet{weber1901, silberstein1907a, silberstein1907b, hestenes1966}; \citet[sec.\ 25]{landaulifshitzfields}; \citet[sec.\ 16.II]{dresden1987}; \citet{riesz1993}.\label{allthesources}}$^,$\footnote{Philosophers of physics have wondered why the magnetic field should be flipped under time reversal and also why quantum wave functions should be complex conjugated under time reversal \citep{albert, callender2000}.  Thinking of the Weber vector as a quantum wave function suggests that these issues may be connected: flipping the sign of $\vec{B}$ is the same operation as complex conjugating $\vec{F}$.}  The other equation \eqref{newmaxwell2} is a restriction on which states of the electromagnetic field are allowed at any instant of time.  If this restriction is satisfied at one time, it will be satisfied at all times.

We will limit our attention in this paper to Maxwell's equations in vacuum, \eqref{newmaxwell1} and \eqref{newmaxwell2}.  Through interactions with charged matter, it would be possible for photons to be created or destroyed.  To avoid this complication and focus on finding a quantum theory of just one photon, we will not consider such interactions with charged matter.

The above wave equation for the Weber vector \eqref{newmaxwell1} can be made to look more similar to the Dirac equation \eqref{thediracequation} if one makes use of the spin-1 matrices $\vec{s}$,\footnote{These matrices appear in a number of the references mentioned in footnote \ref{allthesources}, though note that in \citet{good1957} the definition differs in sign.}
\begin{equation}
s_1=\begin{pmatrix}
0 & 0 & 0 \\
0 & 0 & -i \\
0 & i & 0
\end{pmatrix}\ ,
\quad
s_2=\begin{pmatrix}
0 & 0 & i \\
0 & 0 & 0 \\
-i & 0 & 0
\end{pmatrix}\ ,
\quad
s_3=\begin{pmatrix}
0 & -i & 0 \\
i & 0 & 0 \\
0 & 0 & 0
\end{pmatrix}\ ,
\end{equation}
which obey the commutation relations,
\begin{equation}
[s_i,s_j]=-i \epsilon_{ijk}s_k\ .
\end{equation}
These matrices can be expressed more compactly using the Levi-Civita symbol,
\begin{equation}
(s_i)_{jk}=-i \epsilon_{ijk}
\ .
\label{compactformk}
\end{equation}
This form makes it clear that one can rewrite cross products using these matrices.  For example, the cross product in the wave equation \eqref{newmaxwell1} becomes
\begin{equation}
i \frac{\partial F_j}{\partial t}  = - i c [\vec{s}_{jk}\cdot \del] F_k
\ .
\label{firstform}
\end{equation} 
To align with standard notation for the Dirac equation, the indices on $F$ and $\vec{s}$ will sometimes be dropped.  Whether it is more convenient to write the Weber vector as $F$ (with indices suppressed) or $\vec{F}$ will depend on the equation at hand.  Don't get confused: $F$ and $\vec{F}$ are the same thing.

The wave equation \eqref{firstform} can alternatively be written in terms of the momentum operator $\vec{p}=-i\hbar\del$ as
\begin{equation}
i \hbar \frac{\partial F}{\partial t} = c [\vec{s}\cdot \vec{p}\:] F
\ .
\label{diracform}
\end{equation}
Using $\vec{p}$, the Hamiltonian can be expressed as $\widehat{H}= c\: \vec{s}\cdot \vec{p}$.

Compare \eqref{diracform} (a putative wave equation for the spin-1 massless photon) to the free Dirac equation \eqref{thediracequation} (the wave equation for the spin-$1/2$ massive electron).  Ignoring the mass term (which should not appear in the photon wave equation), the two equations are identical in form.  In the move from \eqref{diracform} to \eqref{thediracequation}, the $\vec{s}$ matrices have been replaced by the $\vec{\alpha}$ matrices and the three-component complex-valued field $F$ has been replaced by the four-component complex-valued field $\psi$.\footnote{One might wonder why the putative higher-spin photon wave function $F$ (or $\phi$ in section \ref{PWFsection}) has fewer components than the lower-spin electron wave function $\psi$.  The reason for this is that $\psi$ has enough degrees of freedom to describe two different spin-$1/2$ particles---the electron and the positron.}  At the level of second-order dynamics, both $F$ and $\psi$ obey the Klein-Gordon equation.  To see that the Weber vector obeys the Klein-Gordon equation, one can apply \eqref{newmaxwell1} twice,
\begin{align}
-\frac{\partial^2 \vec{F}}{\partial t^2}  &= c^2  \: \del \times (\del \times \vec{F})
\nonumber
\\
&=c^2 \left[ \del(\del\cdot\vec{F})-\nabla^2 \vec{F} \right]
\ .
\label{derivingKG}
\end{align}
The $\del\cdot\vec{F}$ term drops out as $\vec{F}$ is divergenceless \eqref{newmaxwell2}, yielding the massless Klein-Gordon equation
\begin{align}
\frac{\partial^2 \vec{F}}{\partial t^2}  = c^2  \nabla^2 \vec{F}
\ .
\label{photonKG}
\end{align}

At this point, we have massaged classical electromagnetism into a form that is beginning to look a lot like quantum mechanics.  But, we're not there yet.  For the Dirac wave function, $\psi^\dagger \psi$ is the probability density and $c \psi^\dagger \vec{\alpha} \psi$ is the probability flux density.  However, $F^{\dagger}F$ is not a probability density.  It is $8\pi$ times the energy density in \eqref{energydensityfield}.  Similarly, $c F^\dagger \vec{s} F$ is not a probability flux density but instead $8\pi$ times the energy flux density in \eqref{poyntingvector}.  One major factor preventing closer alignment here is that the classical electromagnetic field describes large numbers of photons en masse whereas the Dirac wave function $\psi$ describes a single electron.  We need to focus our attention on a single photon.\footnote{One could alternatively seek alignment by introducing a wave function for many electrons (obeying a multi-particle Dirac equation) and comparing this to $F$.  However, such a multi-particle wave function would be defined over configuration space, unlike $F$ which is defined over physical space.}

\section{The Electromagnetic Field of a Single Photon}\label{EMfieldsinglephoton}

\begin{figure}[htb]
\center{\includegraphics[width=13 cm]{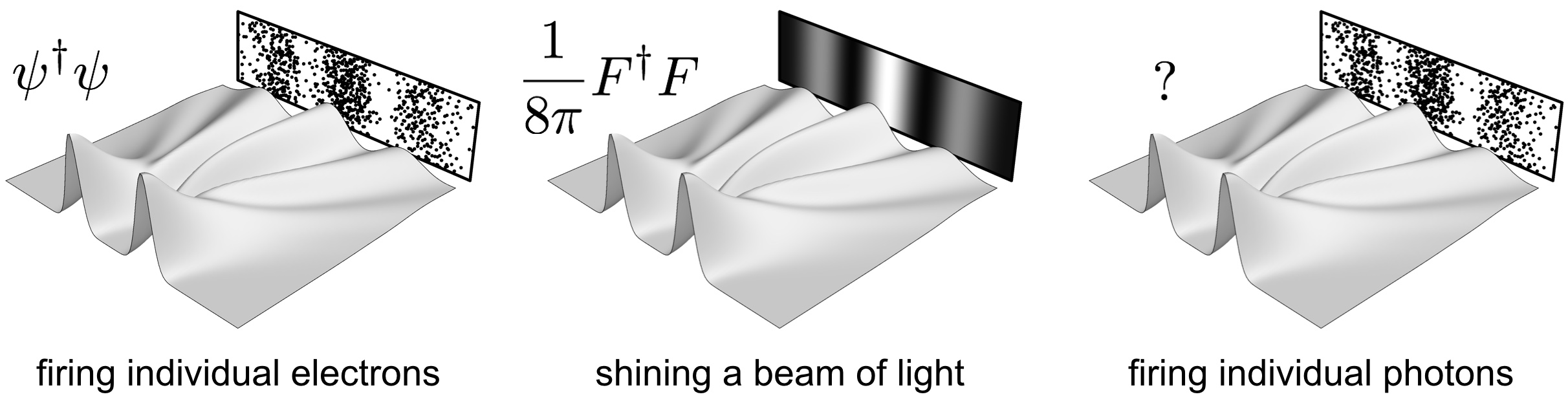}}
\caption{When individual electrons pass through a double-slit setup one-at-a-time, the interference pattern is explained by the electron's quantum wave function.  The evolution of this wave function's amplitude-squared as it approaches the detector is plotted here.  When a beam of light passes through a double-slit setup, the interference pattern is explained by the electromagnetic field.  The field's energy density \eqref{energydensityfield} is plotted here.  When individual photons pass through such a setup, the explanation of the interference pattern is less clear.}
\label{doubleslitfigure}
\end{figure}

In presenting the double-slit experiment for the electron, one describes the interference pattern as generated by the wave-like nature of the electron's wave function and the discrete dots on the screen as having different explanations depending on one's solution to the measurement problem (e.g., collapse events or definite final locations for the electron itself).  In presenting the double-slit experiment for the photon, the explanation of the interference pattern is less obvious (see figure \ref{doubleslitfigure}).  One often starts by considering shining a beam of light upon the slits, i.e., firing many photons at once.  At this point, you can explain the interference pattern without quantum mechanics.  The wave-like nature of the electromagnetic field is sufficient.  If you turn down the intensity of the light source so that one photon is emitted at a time, what explains the interference pattern?  The fact that, when there were many photons, it was the wave-like nature of the electromagnetic field that explained the interference suggests that---as the intensity of light is turned down---this should remain the explanation.  However, comparison to the electron suggests that the wave-like nature of the photon's quantum wave function should be the explanation of the interference.  Which is it?

One could of course respond to this puzzle by picking sides, claiming that it is either the electromagnetic field or the photon wave function which explains the interference pattern.  Alternatively, one could attempt to have it both ways.  Perhaps the electromagnetic field of a single photon \emph{is} its wave function.\footnote{The idea that a weak electromagnetic field might actually be a quantum wave function is provocative and its metaphysical implications are well worth exploring further, as there has been much discussion of the extent to which the quantum wave function resembles a classical field in debates about the ontological status of the wave function (e.g., \citealp{neyalbert}).} One might reasonably expect the laws of electromagnetism to break down when the field becomes sufficiently weak that it only describes a single photon.  But, they do not.  As laws governing the photon's wave function, we will see that they are surprisingly satisfactory (though one may have to make some further modifications to implement their preferred solution to the measurement problem, e.g., by adding another law governing the collapse of the wave function or by introducing a photon that is separate from the electromagnetic field but guided by it in accordance with a guidance equation).

In this section we will examine how weak an electromagnetic field must be in order to plausibly be characterized as describing a single photon.  We need a way of counting how many photons there are in a given state of the electromagnetic field.  (In quantum field theory it is possible to have a state which is in a superposition of there being different numbers of photons.  Let's put this complication aside and press on.\footnote{\citet[pg.\ 540--541, 545]{holland} sees the fact that the number of photons can be indefinite as an obstacle to finding a Bohmian theory in which photons follow definite trajectories (to be contrasted with a Bohmian approach in which the electromagnetic field has a definite configuration).  But, it is possible to have a Bohmian theory in which there is a true number of particles even though the quantum state is in a superposition of different numbers of particles.  See, for example, the way particle creation and annihilation is handled in \citet{dgz2004}.})  We know the energy density of the electromagnetic field \eqref{energydensityfield} and we know from quantum mechanics that the energy of a photon with wave vector $\vec{k}$ is $\hbar k c$ (where $k$ is the wave number, $k=|\vec{k}|$).  But, it's not immediately obvious how to generate a photon number density since we can't divide the energy density by the energy per photon without knowing what energies the photons are supposed to have, i.e., without knowing their wave vectors.  This problem becomes tractable if we Fourier transform our electromagnetic field and work in $\vec{k}$-space,\footnote{See \citet[sec.\ 1.2]{ab1965}.} which we can also call momentum space as the momentum of a photon with wave vector $\vec{k}$ is $\hbar\vec{k}$.

The Fourier transform of an arbitrary vector field $\vec{V}(\vec{x})$ is\footnote{It follows from \eqref{ftransform} that only complex-valued functions $\vec{\widetilde{E}}(\vec{k})$ and $\vec{\widetilde{B}}(\vec{k})$ satisfying $\vec{\widetilde{E}}^*\!\!(\vec{k})=\vec{\widetilde{E}}(-\vec{k})$ and $\vec{\widetilde{B}}^*\!\!(\vec{k})=\vec{\widetilde{B}}(-\vec{k})$ can correspond to real-valued electric and magnetic fields.  Because $\vec{F}(\vec{x})$ is complex-valued, $\vec{\widetilde{F}}(\vec{k})$ does not have such a symmetry.\label{symmetrytrick}}
\begin{align}
\vec{\widetilde{V}}(\vec{k})&=\frac{1}{(2\pi)^{3/2}}\iiint{ \vec{V}(\vec{x}) e^{- i \vec{k} \cdot \vec{x}} \: d^3 x}
\ .
\label{ftransform}
\end{align}
If you Fourier transform the Weber vector, its equation of motion \eqref{newmaxwell1} becomes
\begin{equation}
\frac{\partial \vec{\widetilde{F}}(\vec{k})}{\partial t} =c\:\vec{k}\times\vec{\widetilde{F}}(\vec{k})
\ .
\label{kform}
\end{equation}
The other piece of Maxwell's equations \eqref{newmaxwell2} becomes
\begin{equation}
\vec{k}\cdot\vec{\widetilde{F}}(\vec{k})=0
\ ,
\label{newmaxwell2kspace}
\end{equation}
requiring that, for every wave vector $\vec{k}$, both $\vec{\widetilde{E}}(\vec{k})$ and $\vec{\widetilde{B}}(\vec{k})$ be perpendicular to $\vec{k}$.  In other words, for each wave that is superposed to form the total electromagnetic field, the electric and magnetic fields must be perpendicular to the direction of wave propagation.\footnote{This condition is discussed further in \citet[pg.\ 583--584]{goodnelson1971}.}  Above and below, $\vec{k}$ and $\vec{x}$ arguments have often been written out explicitly to make clear whether we are working in momentum space or ordinary position space. (For now, the dependence on $t$ remains implicit.)

The total energy in the electromagnetic field can either be written as an integral of the energy density in \eqref{energydensityfield} over position space or as an integral of the energy density
\begin{equation}
\widetilde{\rho}^{\,\mathcal{E}}(\vec{k})=\frac{1}{8 \pi}\vec{\widetilde{F}}^*\!\!(\vec{k})\cdot\vec{\widetilde{F}}(\vec{k})
\ 
\label{energydensityfieldkspace}
\end{equation}
in momentum space,
\begin{align}
\mathcal{E}&=\iiint{\rho^{\mathcal{E}}(\vec{x}) \: d^3 x}=\iiint{\widetilde{\rho}^{\,\mathcal{E}}(\vec{k}) \: d^3 k}
\ .
\label{energyintegral}
\end{align}
We can introduce a photon number density in momentum space by dividing the energy density in \eqref{energydensityfieldkspace} by the energy per photon,
\begin{equation}
\widetilde{\rho}^{\,N}\!(\vec{k})=\frac{1}{8 \pi}\frac{\vec{\widetilde{F}}^*\!\!(\vec{k})\cdot\vec{\widetilde{F}}(\vec{k})}{\hbar k c}
\ .
\label{numberdensityfieldkspace}
\end{equation}
This number density can be integrated over momentum space to get the total photon number,\footnote{This expression for total photon number appears in \citet{landau1930, good1957, zeldovich}; \citet[pg.\ 318]{bb1996}; \citet{forcesonfields}.}
\begin{align}
N&=\frac{1}{8 \pi}\iiint{ \frac{\vec{\widetilde{F}}^*\!\!(\vec{k})\cdot\vec{\widetilde{F}}(\vec{k})}{\hbar k c} \: d^3 k}
\nonumber
\\
&=\frac{1}{8 \pi}\iiint{ \frac{\vec{\widetilde{E}}^*\!\!(\vec{k})\cdot\vec{\widetilde{E}}(\vec{k})+\vec{\widetilde{B}}^{*}\!\!(\vec{k})\cdot\vec{\widetilde{B}}(\vec{k})}{\hbar k c} \: d^3 k}
\nonumber
\\
&=\frac{1}{16 \pi^3 \hbar c}\iiint{\!\!\!\iiint{ \frac{\vec{F}(\vec{x})\cdot \vec{F}(\vec{y})}{|\vec{x}-\vec{y}|^2} \: d^3 x\: d^3 y}}
\nonumber
\\
&=\frac{1}{16 \pi^3 \hbar c}\iiint{\!\!\!\iiint{ \frac{\vec{E}(\vec{x})\cdot \vec{E}(\vec{y})+\vec{B}(\vec{x})\cdot \vec{B}(\vec{y})}{|\vec{x}-\vec{y}|^2} \: d^3 x\: d^3 y}}
\ .
\label{fieldnumber}
\end{align}
Here the total photon number has been expressed in four ways: in terms of either the Weber vector or the electric and magnetic fields, in either momentum space or position space.\footnote{The move from the first line of \eqref{fieldnumber} to the second makes use of the symmetry noted in footnote \ref{symmetrytrick}.}

We now have a way to restrict our attention to the electromagnetic field of a single photon.  We can consider only states for which $N$ in \eqref{fieldnumber} is one.  This simple move takes us from classical electromagnetism to a quantum theory of the photon (modulo a solution to the measurement problem).  Seeing the move from classical to quantum physics from this perspective puts emphasis on the old insight of Planck and Einstein that the energy of light is quantized in units of $\hbar k c$.

It makes sense to speak of number densities when there are sufficiently many particles that the number of particles in any small-but-not-too-small region varies relatively smoothly as one scans the space.  However, now that we are considering a single photon, \eqref{numberdensityfieldkspace} cannot be interpreted as a number density.  Instead, it should be interpreted as a probability density (which limits to a number density as the number of photons becomes large),
\begin{equation}
\widetilde{\rho}^{\,p}(\vec{k})=\frac{1}{8 \pi}\frac{\vec{\widetilde{F}}^*\!\!(\vec{k})\cdot\vec{\widetilde{F}}(\vec{k})}{\hbar k c}
\ .
\label{probdensityfieldkspace}
\end{equation}
Although this gives us a way of introducing quantum probabilities into the theory, it does not resolve the problem for regarding the Weber vector as a photon wave function noted at the end of the previous section.  In momentum space, the Weber vector's magnitude squared (when divided by $8 \pi$) gives an energy-weighted probability density---\eqref{numberdensityfieldkspace} times $\hbar k c$---not a straight probability density.  Similarly, we might interpret $\frac{1}{8 \pi}F^{\dagger}F$ (from the end of the previous section) as an energy-weighted probability density in ordinary position space.\footnote{On this picture: Integrating the Weber vector's magnitude squared (over $8 \pi$) in either position or momentum space yields the expectation value of the energy, not the total energy.}  Noting this sort of oddity, some have called $\vec{F}$ the photon \emph{energy} wave function.\footnote{This name comes up in \citet[sec.\ 12.11.5]{mandelwolf}; \citet{bb1996, keller2005}.}

\section{The Photon Wave Function}\label{PWFsection}

At the end of the last section we noted that the Weber vector has a flaw: its square is not a probability density, but instead an energy-weighted probability density (when divided by $8 \pi$).  This is easily remedied.  We must simply remove the energy-weighting.  Following \citet{good1957},\footnote{\citet{good1957} does not restrict his attention to states of the electromagnetic field describing a single photon.  He thus treats $\phi^\dagger \phi$ as a number density, not a probability density---see \eqref{photonprobdensity}.}$^,$\footnote{\citet[pg.\ 191]{pauli}; \citet[eq.\ 1.6]{ab1965}; \citet[pg.\ 637]{mandelwolf} introduce somewhat similar photon wave functions which also involve removing an energy weighting.  Akhiezer and Berestetskii caution against Fourier transforming their momentum-space photon wave function and interpreting the result as a position-space photon wave function (see also \citealp{pauli}).  Their concern is that ``the presence of a photon can be established only as a result of its interaction with charges'' and that the force a charge will feel at a given point is determined by the electric and magnetic fields at that point.  Since a wave function like $\phi$ at a point $\vec{x}$ is dependent on the electric and magnetic fields at distant points---see \eqref{expandingphi}---they argue that it will not give the right interactions.  They conclude that ``the concept of probability density for the localization of a photon does not exist.''  This argument is a bad mix of classical and quantum ideas.  The interaction is modeled classically and yet quantum probabilities are expected to emerge.  When the electromagnetic field of a single photon hits the screen at the end of a double-slit experiment like the one depicted in figure \ref{doubleslitfigure}, its interaction with the screen is not a matter of weak fields exerting weak forces all over the screen.  The photon is found at just one location.  Via \eqref{photonprobdensity}, the position-space photon wave function tells us how likely each location is.} we can introduce a photon wave function $\vec{\phi}$ related to the Weber vector in momentum space by
\begin{equation}
\vec{\widetilde{\phi}}(\vec{k})=\frac{\vec{\widetilde{F}}(\vec{k})}{\sqrt{8\pi\hbar k c}}
\ ,
\label{defofphi}
\end{equation}
where the factor of $\sqrt{\hbar k c}$ counteracts the energy-weighting and the factor of $\sqrt{8\pi}$ corrects the above-mentioned factor of $8 \pi$.  In position space, the photon wave function is
\begin{align}
\vec{\phi}(\vec{x})&=\frac{1}{\sqrt{8\pi}}\frac{1}{(2\pi)^{3/2}}\iiint{ \frac{\vec{\widetilde{F}}(\vec{k}) e^{i \vec{k}\cdot\vec{x}}}{\sqrt{\hbar k c}} \: d^3 k }
\nonumber
\\
&=\frac{1}{\sqrt{8\pi}}\frac{1}{(2\pi)^3}\iiint{ \left[ \frac{e^{i \vec{k}\cdot\vec{x}}}{\sqrt{\hbar k c}} \iiint{\left(\vec{E}(\vec{y})+i\vec{B}(\vec{y})\right)e^{-i \vec{k}\cdot\vec{y}} \: d^3 y }\right] d^3 k}
\label{expandingphi}
\ .
\end{align}
The second line shows how the photon wave function is constructed from the electric and magnetic fields.  The value of $\vec{\phi}$ at a point $\vec{x}$ is not determined solely by the electric and magnetic fields at that point.  It depends on the fields elsewhere as well.\footnote{Because $\vec{\phi}$ at $\vec{x}$ is not fixed by $\vec{E}$ and $\vec{B}$ at $\vec{x}$, one might conclude that $\vec{\phi}$ is not a ``local'' field.  If the electromagnetic field is fundamental and the photon wave function is defined from it, I think it would be reasonable to classify $\vec{\phi}$ as non-local.  However, one could invert \eqref{expandingphi} and express the electromagnetic field non-locally in terms of the photon wave function.  If we take the wave function to be fundamental and the electromagnetic field to be defined from it, then it is the electromagnetic field which should be deemed non-local.  Not knowing how to settle such a question of fundamentality at this point, I think it is best to just keep in mind that switching from one field to the other is not a local maneuver.}

The photon wave function obeys the same wave equation as the Weber vector, which can be expressed in the form of \eqref{newmaxwell1}, \eqref{diracform}, or \eqref{kform}:
\begin{align}
i\frac{\partial \vec{\phi}}{\partial t}  &= c\del\times\vec{\phi}
\nonumber
\\
i\hbar\frac{\partial \phi}{\partial t}  &= c \: \vec{s}\cdot \vec{p}\: \phi
\nonumber
\\
\frac{\partial \vec{\widetilde{\phi}}}{\partial t} &=c\:\vec{k}\times\vec{\widetilde{\phi}}
\ .
\label{photonschrodinger}
\end{align}
The fact that $\vec{\phi}$ obeys the above wave equation follows from \eqref{expandingphi} and Maxwell's equations.  As the Weber vector must be divergenceless \eqref{newmaxwell2}, $\vec{\phi}$ must as well
\begin{align}
\del\cdot\vec{\phi}=0
\ .
\label{newmaxwell2phi}
\end{align}
As in \eqref{newmaxwell2kspace}, this means that for every wave vector $\vec{k}$, the wave function amplitude $\vec{\widetilde{\phi}}(\vec{k})$ must be perpendicular to $\vec{k}$.  From \eqref{photonschrodinger} and \eqref{newmaxwell2phi}, it follows that that the second-order dynamics of $\vec{\phi}$ are given by the massless Klein-Gordon equation \eqref{photonKG}.  In the above equations and below, I have taken the same liberty of notation with $\vec{\phi}$ as I did with $\vec{F}$ (explained in section \ref{RSsection}): $\phi$ and $\vec{\phi}$ are different notations for the same thing.

The probability density in position space is
\begin{equation}
\rho^{\,p}=\phi^\dagger \phi
\ 
\label{photonprobdensity}
\end{equation}
and the probability flux density is
\begin{equation}
\vec{J}^{\:p}=c \phi^\dagger \vec{s} \phi
\ .
\label{photoncurrentdensity}
\end{equation}
These quantities take exactly the same from as the probability density $\psi^\dagger \psi$ and probability flux density $c \psi^\dagger \vec{\alpha} \psi$ for the Dirac wave function (see table \ref{table1}).  The fact that \eqref{photonprobdensity} and \eqref{photoncurrentdensity} obey a continuity equation of the form $\frac{\partial \rho^{\,p}}{\partial t}=-\del\cdot\vec{J}^{\:p}$ follows directly from \eqref{photonschrodinger}.  Note that, because of the way $\phi$ is related to $\vec{E}$ and $\vec{B}$ \eqref{expandingphi}, it is possible for the probability density to be non-zero where $\vec{E}$ and $\vec{B}$ are zero.

\begin{table}[h!]
\centering
\caption{The proposed quantum theory of the photon, derived as a rewriting of classical electromagnetism, closely parallels the standard quantum theory of the electron (where the electron's wave function evolves by the Dirac equation).}
\begin{align}
\begin{array}{rll}
 &\mbox{\sc \underline{The Photon}} & \mbox{\sc \underline{The Electron}}
\nonumber
\vspace*{6 pt}
\\
\mbox{Wave Function}\quad&\phi(\vec{x},t)&\psi(\vec{x},t)
\vspace*{2 pt}
\nonumber
\\
\mbox{Number of Components}\quad&3&4
\nonumber
\\
\mbox{Wave Equation}\quad&i\hbar\frac{\partial \phi}{\partial t}  = c \: \vec{s}\cdot \vec{p}\: \phi \quad&i\hbar \frac{\partial \psi}{\partial t}=\left(c\:\vec{\alpha}\cdot\vec{p}+\beta m c^2 \right)\psi
\vspace*{2 pt}
\nonumber
\\
\mbox{Probability Density}\quad&\phi^\dagger \phi&\psi^\dagger \psi
\vspace*{2 pt}
\nonumber
\\
\mbox{Probability Current}\quad&c \phi^\dagger \vec{s} \phi&c \psi^\dagger \vec{\alpha} \psi
\end{array}
\nonumber
\end{align}
\label{table1}
\end{table}

If we had instead taken the Weber vector to be the wave function of the photon, it would have been natural to take the probability density to be proportional to $F^\dagger F$ instead of $\phi^\dagger \phi$\footnote{The idea that probability density is proportional to $F^\dagger F$ has been proposed by \citet{archibald1955, wesley1984, esposito1999}.}---making the probability density proportional to the energy density \eqref{energydensityfield}---
\begin{equation}
\rho^{\,p}\stackrel{\bm{?}}{\propto}\rho^{\mathcal{E}}=\frac{1}{8\pi} F^\dagger F =\frac{1}{8 \pi}\vec{F}^{*}\!\cdot\vec{F}
\ .
\label{badphotonprobability}
\end{equation}
The idea here is that the energy density can be divided by some constant with units of energy to get an appropriately normalized probability density.  In the same spirit, one could take the probability current to be proportional to the Poynting vector \eqref{poyntingvector} (dividing by the same constant),
\begin{equation}
\vec{J}^{\:p}\stackrel{\bm{?}}{\propto}\vec{S}=\frac{c}{8 \pi} F^\dagger \vec{s} F =\frac{c}{8 \pi i}\vec{F}^* \! \times \! \vec{F}
\ .
\label{badphotoncurrent}
\end{equation}
I don't think that the expression for the probability density in \eqref{badphotonprobability} is particularly well-motivated since we are simply dividing the energy density by a fixed constant even though we know that a photon's energy is dependent on its wave number---a dependence which is explicitly taken into account in \eqref{defofphi}.  But, because energy is conserved, the above expressions are not completely absurd; they will respect conservation of probability.  In the following section, we will compare the probability density and probability current proposed earlier, \eqref{photonprobdensity} and \eqref{photoncurrentdensity}, to these alternative expressions derived from the Weber vector.

Let us now close this section by revisiting the project of finding a Bohmian theory in which there is a photon particle distinct from the photon wave function.  If we take the particle to be guided by the photon wave function proposed here, it would be natural to introduce a guidance equation according to which the photon velocity is equal to the probability flux density at the photon's location \eqref{photonprobdensity} divided by the probability density at that location \eqref{photoncurrentdensity},
\begin{equation}
\vec{v}=\frac{\vec{J}^{\:p}}{\rho^{\:p}}=\frac{c \phi^\dagger \vec{s} \phi}{\phi^\dagger \phi} =\frac{-i c\:\vec{\phi}^* \! \times \! \vec{\phi}}{\vec{\phi}^* \cdot \vec{\phi}}
\ ,
\label{photonguidance}
\end{equation}
just as the Bohmian electron velocity\footnote{See \citet{bohm1953}; \citet[sec.\ 12.2]{bohmhiley}; \citet[sec.\ 12.2]{holland}.} is the Dirac probability flux density divided by the probability density,
\begin{equation}
\vec{v}=\frac{c \psi^\dagger \vec{\alpha} \psi}{\psi^\dagger \psi}
\ .
\label{electronguidance}
\end{equation}
This guidance equation for the photon \eqref{photonguidance} does not require the photon to always travel at the speed of light.  However, it does ensure that (as for the electron) the velocity of the photon cannot exceed the speed of light.  The velocity is maximized at $c$ when the real and imaginary parts of the photon wave function are perpendicular and equal in magnitude.  Using \eqref{expandingphi}, the velocity of the photon can be expressed in terms of the electromagnetic fields.  Because $\phi$ at a point $\vec{x}$ depends on the electric and magnetic fields away from $\vec{x}$, the velocity of the photon will not be determined by the values of the electric and magnetic fields at its location.

One arrives at a different guidance equation for the photon if one takes the Weber vector to be the wave function of the photon.  One can divide the probability current in \eqref{badphotoncurrent} by the probability density in \eqref{badphotonprobability} to get a particle velocity
\begin{equation}
\vec{v}=\frac{\vec{S}}{\rho^{\mathcal{E}}}=\frac{c F^\dagger \vec{s} F}{F^\dagger F} =\frac{-i c\:\vec{F}^* \! \times \! \vec{F}}{\vec{F}^* \cdot \vec{F}}
\ .
\label{badphotonguidance}
\end{equation}
This photon velocity has been defended by \citet{wesley1984} and criticized by \citet[sec.\ 11.2]{bohmhiley}; \citet[sec. 12.6.2]{holland}.  As far as I am aware, the photon guidance equation in \eqref{photonguidance} has not been proposed elsewhere.

\section{Lorentz Transformations}\label{EXsection}

Because the Weber vector $\vec{F}$ obeys Maxwell's equations in every inertial frame, the photon wave function $\vec{\phi}$ constructed from $\vec{F}$ via \eqref{expandingphi} in each frame will always obey its wave equation \eqref{photonschrodinger}.  The wave equation is a relativistic wave equation.  Problems with relativity arise in the probabilities.

The probability density and probability current for the Dirac wave function together form a four-current, i.e., they together transform as a four-vector under Lorentz transformations.\footnote{See, e.g., \citet[sec.\ 1.3 and 2.2]{bjorkendrell}; \citet[sec.\ 4c]{schweberQFT}.}  In this section we will see that the photon probability density and probability current density proposed above do not always transform as a four-vector, though they do transform in that way more often than you might have expected.  Because the probabilities do not transform properly, two moving observers who agree on the photon wave function may radically disagree on the predicted results of experiments.  This problem with Lorentz transformations cannot be solved by suggesting that the Weber vector is the correct wave function for the photon and the probability density is proportional to $F^\dagger F$, not $\phi^\dagger \phi$---as in \eqref{badphotonprobability} and \eqref{badphotoncurrent}.\footnote{\citet{kiessling2017} have recognized that the probability density and probability current defined from the Weber vector, \eqref{badphotonprobability} and \eqref{badphotoncurrent}, do not transform together as a four-vector and, because of this defect, concluded that the Weber vector is unacceptable as a wave function for the photon.}  In that case, probability density would be proportional to energy density.  But, the energy and momentum of the electromagnetic field do not together transform as a four-vector.\footnote{The fact that the energy and momentum of the electromagnetic field do not form a four-vector is well-known.  The energy and momentum form a four-by-four tensor (the energy-momentum tensor) when combined with the momentum flux density of the field.}  Neither $\vec{\phi}$ nor $\vec{F}$ yield probabilities which transform properly.  Thus, neither is acceptably relativistic to serve as a relativistic wave function for the photon.\footnote{At least, neither is acceptably relativistic when paired with its natural probability density and current: \eqref{photonprobdensity} and \eqref{photoncurrentdensity} or \eqref{badphotonprobability} and \eqref{badphotoncurrent}.}

We will first examine the way a single circularly polarized plane wave appears to multiple moving observers.  Here the probabilities generated from the photon wave function transform properly but those generated from the Weber vector do not.  Then, we will move to the more complex case of two oppositely oriented plane waves.  For boosts along the direction of wave propagation, the probabilities generated from the photon wave function transform properly.  However, for boosts perpendicular to the direction of wave propagation the probabilities do not transform properly.  An interference pattern emerges that would not have been predicted from the initial probability density and probability current.  At the end of the section, I will discuss the bearing of these examples on the project of defining a Bohmian guidance equation for the photon.

\begin{figure}[h!]
\center{\includegraphics[width=13 cm]{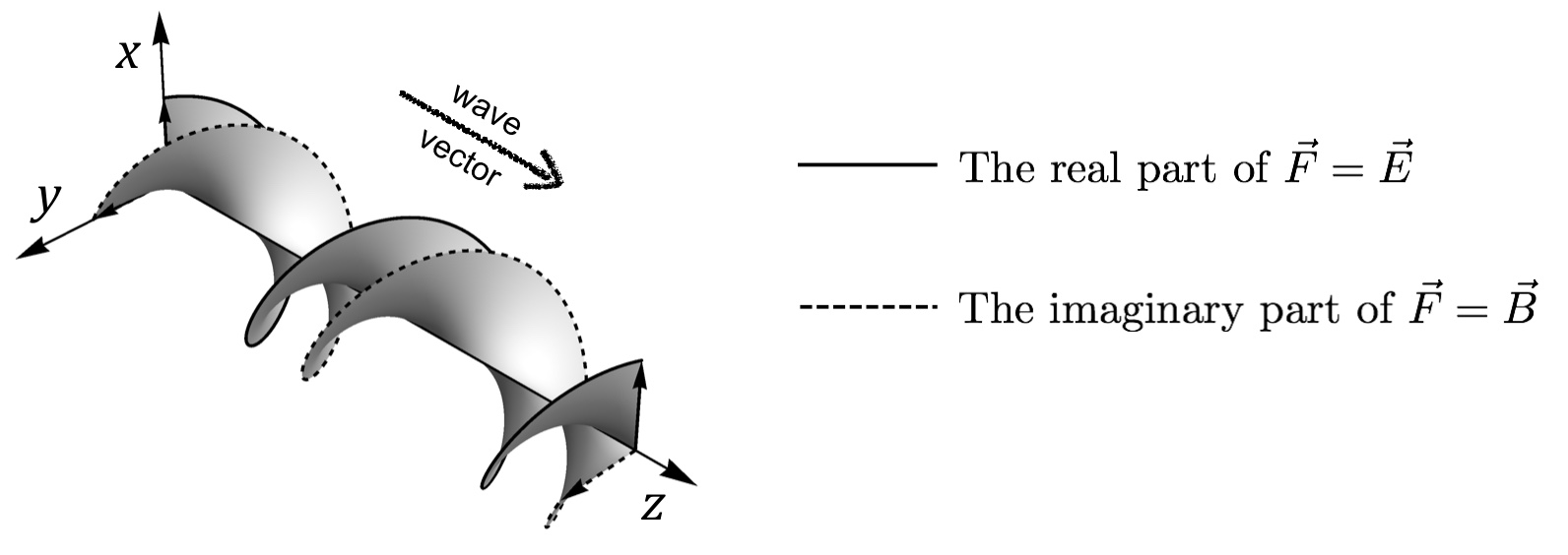}}
\caption{This figure depicts the right-moving circularly polarized wave in \eqref{simpleplanewavefield}.  The $z$-axis indicates location in space whereas the $x$ and $y$ axes are used to isolate components of vectors.  The solid arrow represents the real part of the Weber vector $\vec{F}$, the electric field $\vec{E}$.  The dotted arrow represents the imaginary part of the Weber vector $\vec{F}$, the magnetic field $\vec{B}$.  Each of these arrows traces out a curve as one moves along the $z$-axis.}
\label{planewavefigure}
\end{figure}

\begin{figure}[h!]
\center{\includegraphics[width=13 cm]{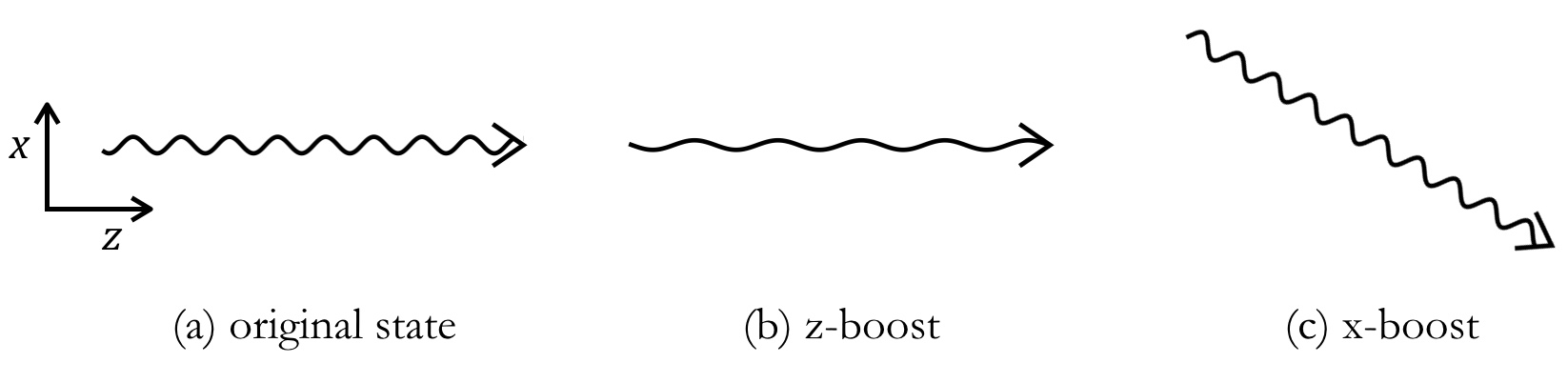}}
\caption{Part (a) of this figure simplifies the detailed depiction of \eqref{simpleplanewavefield} in figure \ref{planewavefigure}.  Part (b) shows how this state of the electromagnetic field would appear to an observer moving in the $z$-direction \eqref{WeberLTofplanewave}, displaying the change in frequency and intensity assuming the observer is moving at half the speed of light.  Part (c) shows the perspective of an observer moving in the $x$-direction \eqref{xboostedstatevectors}, who will see the wave as propagating in a different direction within their frame of reference---moving partly backwards as the observer moves forwards.}
\label{singlewavetransformations}
\end{figure}

Consider a right-handed circularly polarized plane wave of intensity $I$ propagating to the right---in the positive $z$-direction---with wave number $k_R$.  Suppose the electromagnetic field as a function of space and time is
\begin{equation}
\vec{F}(\vec{x},t) = \sqrt{\frac{4 \pi I}{c}} e^{i k_R (z-ct)}
\begin{pmatrix}
1 \\
i \\
0
\end{pmatrix}
\ ,
\label{simpleplanewavefield}
\end{equation}
depicted carefully in figure \ref{planewavefigure} and more impressionistically in figure \ref{singlewavetransformations}.a.  This state has constant energy density $\frac{I}{c}$ and energy flux density $I\hat{z}$. In momentum space, this is a delta function state localized around $\vec{k}=(0,0,k_R)$.  Using \eqref{expandingphi}, it is immediately clear that the photon wave function $\vec{\phi}$ has the same form as the Weber vector $\vec{F}$,
\begin{equation}
\vec{\phi}(\vec{x},t) = \sqrt{\frac{I}{2 \hbar k_R c^2}} e^{i k_R (z-ct)}
\begin{pmatrix}
1 \\
i \\
0
\end{pmatrix}
\ .
\label{simpleplanewavestate}
\end{equation}
The probability density and current density for this state are
\begin{align}
\rho^{\,p}(\vec{x},t)&=\frac{I}{\hbar k_R c^2}
\nonumber
\\
\vec{J}^{\:p}(\vec{x},t)&=\frac{I}{\hbar k_R c}\hat{z}
\ .
\label{probflowplanewave}
\end{align}
Probability flows in the direction of wave propagation with constant flux density.  As one would expect for a quantum state describing a particle with definite momentum, this state is not normalizable (as the probability density is uniform throughout space).\footnote{Of course, this pathology could be easily remedied by imagining space to be finite and imposing periodic boundary conditions.}

Next, let us consider what this plane wave would look like in the frame of reference of an observer moving with velocity $u$ in the $z$-direction, the direction in which the wave is propagating.\footnote{For derivation of the way electromagnetic waves transform under Lorentz boosts, see discussions of the relativistic doppler effect in, e.g., \citet[sec.\ 7]{einstein1905}; \citet[prob.\ 12.47]{griffiths}; \citet[pg.\ 467--468]{pollackstump}.  Alternatively, one can calculate the way these waves transform using the general transformation properties of the Weber vector (\citealp[eq.\ 25.5]{landaulifshitzfields}; \citealp[sec.\ 35]{goodnelson1971}).}  In this frame, the Weber vector becomes
\begin{align}
\vec{F}'(\vec{x}',t') &= \sqrt{\frac{4 \pi I'}{c}} e^{i k_R' (z'-ct')}
\begin{pmatrix}
1 \\
i \\
0
\end{pmatrix}
= \sqrt{\frac{1-\frac{u}{c}}{1+\frac{u}{c}}}\sqrt{\frac{4 \pi I}{c}} e^{i k_R' (z'-ct')}
\begin{pmatrix}
1 \\
i \\
0
\end{pmatrix}
\label{WeberLTofplanewave}
\ 
\end{align}
and the photon wave function derived from it via \eqref{expandingphi} becomes
\begin{align}
\vec{\phi}'(\vec{x}',t') &= \sqrt{\frac{I'}{2 \hbar k_R' c^2}} e^{i k_R' (z'-ct')}\begin{pmatrix}
1 \\
i \\
0
\end{pmatrix}
= \left(\frac{1-\frac{u}{c}}{1+\frac{u}{c}}\right)^{1/4}\sqrt{\frac{I}{2 \hbar k_R c^2}} e^{i k_R' (z'-ct')}
\begin{pmatrix}
1 \\
i \\
0
\end{pmatrix}
\ ,
\label{phiLTofplanewave}
\end{align}
where $k_R'$ is the new wave number,
\begin{equation}
k_R'=\sqrt{\frac{1-\frac{u}{c}}{1+\frac{u}{c}}}k_R
\ ,
\label{kRLT}
\end{equation}
and the wave's intensity is now
\begin{equation}
I'=\frac{1-\frac{u}{c}}{1+\frac{u}{c}}I
\ .
\end{equation}
The wave number has decreased and the wavelength increased as the wave has been redshifted (depicted in figure \ref{singlewavetransformations}.b).  The probability density and current derived from this state via \eqref{photonprobdensity} and \eqref{photoncurrentdensity} agree with what one would expect from \eqref{probflowplanewave} if the density and current together transform as a four-vector,
\begin{align}
\rho^{\,p}{}'(\vec{x}',t')&=\sqrt{\frac{1-\frac{u}{c}}{1+\frac{u}{c}}}\frac{I}{\hbar k_R c^2}=\gamma \rho^{\,p}(\vec{x},t)-\gamma \frac{u}{c^2} J_3^{\:p}(\vec{x},t)
\nonumber
\\
J_1^{\:p}{}'(\vec{x}',t')&=J_2^{\:p}{}'(\vec{x}',t')=0
\nonumber
\\
J_3^{\:p}{}'(\vec{x}',t')&=\sqrt{\frac{1-\frac{u}{c}}{1+\frac{u}{c}}}\frac{I}{\hbar k_R c}= - \gamma u \rho^{\,p}(\vec{x},t)+\gamma J_3^{\:p}(\vec{x},t)
\ ,
\label{zboostedprobflow}
\end{align}
where $\gamma=1/\text{\footnotesize $\sqrt{1-\frac{u^2}{c^2}}$}$.  If we were instead to use the alternative probability density and probability current derived from the Weber vector in \eqref{badphotonprobability} and \eqref{badphotoncurrent}, we would not see the probability density and current transforming as a four-vector in this case (because the energy and momentum densities of the electromagnetic field do not transform as a four-vector).  This alternative proposal thus fails right out of the gate.  The probability density and current in \eqref{photonprobdensity} and \eqref{photoncurrentdensity} do much better, transforming properly in a number of non-trivial cases.  However, we will see that they do not always transform properly.

Now, let us ask how \eqref{simpleplanewavefield} would appear to an an observer moving with velocity $u$ in the $x$-direction, perpendicular to the direction in which the wave is propagating.  To such an observer, the wave vector has a component in the $z$-direction (as it was moving in this direction in the original frame) as well as a component in the minus $x$-direction (as the observer is moving in the $x$-direction relative to the initial frame),
\begin{equation}
\vec{k}_R'=\begin{pmatrix}
- \frac{\gamma u}{c} \\
0 \\
1
\end{pmatrix} k_R
\ .
\label{xboostedwavevector}
\end{equation}
The new intensity is
\begin{equation}
I'=\gamma^2 I
\ .
\end{equation}
Rotating the electric and magnetic fields to be perpendicular to the new wave vector, the transformed state is
\begin{align}
\vec{F}'(\vec{x}',t')&=\gamma \sqrt{\frac{4 \pi I}{c}} e^{i \vec{k}_R' \cdot \vec{x}'-i k_R' c t'}
\begin{pmatrix}
\frac{1}{\gamma}\\
i \\
\frac{u}{c}
\end{pmatrix}
\nonumber
\\
\vec{\phi}'(\vec{x}',t')&=\sqrt{\gamma} \sqrt{\frac{I}{2 \hbar k_R c^2}} e^{i \vec{k}_R' \cdot \vec{x}'-i k_R' c t'}
\begin{pmatrix}
\frac{1}{\gamma} \\
i \\
\frac{u}{c}
\end{pmatrix}
\ ,
\label{xboostedstatevectors}
\end{align}
depicted in figure \ref{singlewavetransformations}.c.  The probability density and current for this state are
\begin{align}
\rho^{\,p}{}'(\vec{x}',t')&=\gamma \frac{I}{\hbar k_R c^2}
\nonumber
\\
\vec{J}^{\:p}{}'(\vec{x}',t')&=\frac{I}{\hbar k_R c}\begin{pmatrix}
- \frac{\gamma u}{c} \\
0 \\
1
\end{pmatrix}
\ ,
\label{probflowxboosted}
\end{align}
just as one would expect from transforming \eqref{probflowplanewave} together as a four-vector.  The probability density has picked up a factor of $\gamma$ due to length contraction and the probability current has picked up an $x$-component so that it remains aligned with the wave vector \eqref{xboostedwavevector}.

\begin{figure}[h!]
\center{\includegraphics[width=13 cm]{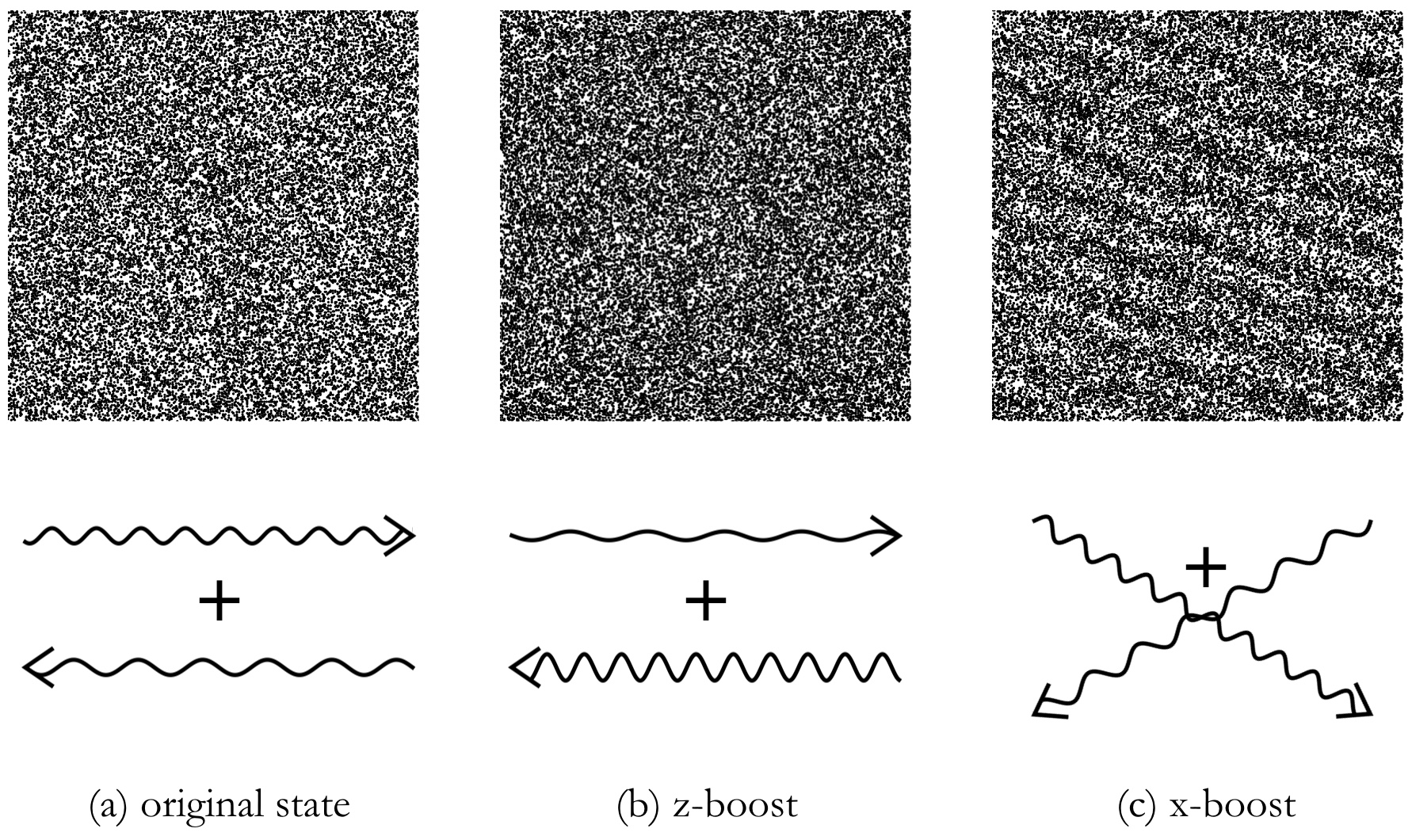}}
\caption{The wavy arrows depict the state of the electromagnetic field in \eqref{twowavestate} as seen in the original frame and by observers moving in the $z$ and $x$-directions (as in figure \ref{singlewavetransformations}).  Here we've assumed the left-moving wave has a smaller wave number than the right-moving wave.  The top squares display the expected patterns of position observations derived from \eqref{photonprobdensity} in \eqref{probflowsuperposition}, \eqref{zboostedprobflowsuperposition}, and \eqref{probflowsuperpositionxboosted}.  The first two plots show points that are randomly distributed, more densely in the second than the first.  The third plot shows a faint interference pattern, which would get sharper as one increased the velocity of the moving observer.  This interference pattern would not be expected if the probability density and current in \eqref{probflowsuperposition} transformed the way relativistic probability densities and currents should.  The presence of this interference pattern shows that the quantum theory of the photon we have been examining is not an acceptable relativistic quantum theory.}
\label{doublewavetransformations}
\end{figure}

Let's move to a somewhat more complicated scenario.  Consider a superposition of two circularly polarized waves with equal intensity $I$ propagating in opposite directions along the $z$-axis,
\begin{equation}
\vec{F}(\vec{x},t) = \sqrt{\frac{4 \pi I}{c}}\left[
e^{i k_R (z-ct)}
\begin{pmatrix}
1 \\
i \\
0
\end{pmatrix}
+
e^{i k_L (z+ct)}
\begin{pmatrix}
1 \\
-i \\
0
\end{pmatrix}
\right]
\ ,
\label{twowavefield}
\end{equation}
depicted in figure \ref{doublewavetransformations}.a.  Here the right-moving wave has wave number $k_R$ and the left-moving wave has wave number $k_L$.  This state has a constant energy density $\frac{2I}{c}$ and no energy flux density (because the two waves have equal intensity and opposite orientations).  The photon wave function one generates from $\vec{F}$ via \eqref{expandingphi} is
\begin{equation}
\vec{\phi}(\vec{x},t) = \sqrt{\frac{I}{2 \hbar c^2}}\left[
\frac{e^{i k_R (z-ct)}}{\sqrt{k_R}}
\begin{pmatrix}
1 \\
i \\
0
\end{pmatrix}
+
\frac{e^{i k_L (z+ct)}}{\sqrt{k_L}}
\begin{pmatrix}
1 \\
-i \\
0
\end{pmatrix}
\right]
\ .
\label{twowavestate}
\end{equation}
The probability density and probability current for this state are
\begin{align}
\rho^{\,p}(\vec{x},t)&=\frac{I}{\hbar c^2}\left(\frac{1}{k_R}+\frac{1}{k_L}\right)
\nonumber
\\
\vec{J}^{\:p}(\vec{x},t)&=\frac{I}{\hbar c}\left(\frac{1}{k_R}-\frac{1}{k_L}\right)\hat{z}
\ ,
\label{probflowsuperposition}
\end{align}
independent of location and time.  Though there is no flow of energy, there is a flow of probability in the direction of the wave with the smaller wave number.  Because the left-moving wave corresponds to a photon energy of $\hbar k_L c$ and the right-moving wave to a photon energy of $\hbar k_R c$, the flow of energy is not proportional to the flow of probability, as in \eqref{probflowplanewave}.

As described by an observer moving with velocity $u$ in the $z$-direction, \eqref{twowavestate} becomes
\begin{align}
\vec{F}'(\vec{x}',t') &= \sqrt{\frac{4 \pi I}{c}}\left[
\sqrt{\frac{1-\frac{u}{c}}{1+\frac{u}{c}}}e^{i k_R' (z'-ct')}
\begin{pmatrix}
1 \\
i \\
0
\end{pmatrix}
+
\sqrt{\frac{1+\frac{u}{c}}{1-\frac{u}{c}}}e^{i k_L' (z'+ct')}
\begin{pmatrix}
1 \\
-i \\
0
\end{pmatrix}
\right]
\nonumber
\\
\vec{\phi}'(\vec{x}',t') &=\sqrt{\frac{I}{2 \hbar c^2}}\left[
 \left(\frac{1-\frac{u}{c}}{1+\frac{u}{c}}\right)^{1/4}\frac{e^{i k_R' (z'-ct')}}{\sqrt{k_R}}
\begin{pmatrix}
1 \\
i \\
0
\end{pmatrix}
+
 \left(\frac{1+\frac{u}{c}}{1-\frac{u}{c}}\right)^{1/4}\frac{e^{i k_L' (z'+ct')}}{\sqrt{k_L}}
\begin{pmatrix}
1 \\
-i \\
0
\end{pmatrix}
\right]
\ ,
\label{twowaveLT}
\end{align}
where the right-moving wave is redshifted, as in \eqref{kRLT}, and the left-moving wave is blueshifted,
\begin{align}
k_L'&=\sqrt{\frac{1+\frac{u}{c}}{1-\frac{u}{c}}}k_L
\ .
\end{align}
The new probability density and current are
\begin{align}
\rho^{\,p}{}'(\vec{x}',t') &=\frac{I}{\hbar c^2}\left(\sqrt{\frac{1-\frac{u}{c}}{1+\frac{u}{c}}}\frac{1}{k_R}+\sqrt{\frac{1+\frac{u}{c}}{1-\frac{u}{c}}}\frac{1}{k_L}\right)
\nonumber
\\
\vec{J}^{\:p}{}'(\vec{x}',t') &=\frac{I}{\hbar c}\left(\sqrt{\frac{1-\frac{u}{c}}{1+\frac{u}{c}}}\frac{1}{k_R}-\sqrt{\frac{1+\frac{u}{c}}{1-\frac{u}{c}}}\frac{1}{k_L}\right)\hat{z}
\ ,
\label{zboostedprobflowsuperposition}
\end{align}
agreeing with the result one would get by transforming the density and current in \eqref{probflowsuperposition} as a four-vector, as in \eqref{zboostedprobflow}.  Thus far, the probabilities have transformed properly.  In the next case, they do not.

As our final case, let us consider the superposed state \eqref{twowavefield} from the perspective of an observer moving with velocity $u$ in the $x$-direction.  The photon wave function becomes
\begin{align}
\vec{\phi}'(\vec{x}', t') = 
\sqrt{\gamma}
\sqrt{\frac{I}{2 \hbar c^2}}\left[
\frac{e^{i \vec{k}_R' \cdot \vec{x}'-i k_R' c t'}}{\sqrt{k_R}}
\begin{pmatrix}
\frac{1}{\gamma} \\
i \\
\frac{u}{c}
\end{pmatrix}
+
\frac{e^{-i \vec{k}_L' \cdot \vec{x}'+i k_L' c t'}}{\sqrt{k_L}}
\begin{pmatrix}
\frac{1}{\gamma} \\
-i \\
-\frac{u}{c}
\end{pmatrix}
\right]
\ ,
\end{align}
where $\vec{k}_R'$ is as in \eqref{xboostedwavevector} and $\vec{k}_L'$ is similarly
\begin{equation}
\vec{k}_L'=\begin{pmatrix}
- \frac{\gamma u}{c} \\
0 \\
-1
\end{pmatrix} k_L
\ .
\end{equation}
The two waves that were moving directly towards one another in the initial state are now crossing at an angle (see figure \ref{doublewavetransformations}.c).  This transformed state has probability density and current
\begin{align}
\rho^{\,p}{}'(\vec{x}',t') &=\gamma\frac{I}{\hbar c^2}\left\{ \frac{1}{k_R} - \frac{2}{\sqrt{k_R k_L}}\left(\frac{u^2}{c^2}\right)\cos\left[( \vec{k}_L' +\vec{k}_R' )\cdot \vec{x}'- ( k_L' + k_R' ) c t' \right]+\frac{1}{k_L} \right\}
\nonumber
\\
\vec{J}^{\:p}{}'(\vec{x}',t') &=\frac{I}{\hbar c}
\begin{pmatrix}
 \frac{-\gamma u}{c} \left\{\frac{1}{k_R}  - \frac{2}{\sqrt{k_R k_L}} \cos\left[( \vec{k}_L' +\vec{k}_R' )\cdot \vec{x}'- ( k_L' + k_R' ) c t' \right] + \frac{1}{k_L}\right\}\\
0 \\
\frac{1}{k_R}  - \frac{2}{\sqrt{k_R k_L}}
\left( \frac{u}{c} \right) \sin\left[( \vec{k}_L' +\vec{k}_R' )\cdot \vec{x}'- ( k_L' + k_R' ) c t' \right]- \frac{1}{k_L}
\end{pmatrix}
\ .
\label{probflowsuperpositionxboosted}
\end{align}
This result serves as a counterexample to the thought that the probability density and current might transform like a four-vector.  Neither the probability density nor the current match the result you would get by transforming \eqref{probflowsuperposition} as a four-vector.  The probability density in \eqref{probflowsuperposition} was uniform in space and time.  Its Lorentz transformation should be uniform as well---just picking up a factor of $\gamma$ from length contraction, as in \eqref{probflowxboosted}.  However, the probability density in \eqref{probflowsuperpositionxboosted} is not uniform.  It displays an interference pattern as the two waves cross one another (shown in figure \ref{doublewavetransformations}.c).

This un-four-vector-like transformation is not acceptable for a relativistic quantum theory of the photon.  An observer in the initial frame and an observer moving at some velocity in the $x$-direction will disagree on the expected results of repeated position measurements for photons in the state \eqref{twowavestate}.  Using the probability density in \eqref{probflowsuperposition}, the original observer would expect to see a uniform distribution of detected positions and (thinking relativistically about how the same events would appear to a moving observer) would expect the moving observer to similarly see a uniform distribution, though a denser distribution due to length contraction---as in \eqref{probflowxboosted}.  By contrast, the moving observer would expect to see an oscillating density of detected positions in accord with \eqref{probflowsuperpositionxboosted}, and, by the lights of special relativity, would expect those same events to appear as an oscillating density of detected positions to the original observer.  This sort of disagreement cannot be tolerated in a relativistic theory.  We have developed an interesting and consistent quantum theory of the photon, just not a relativistic one.\footnote{It is interesting to think about where problem with relativity arises if one approaches the situation from the perspective of the many-worlds interpretation.  The basic physics just includes a wave function obeying the wave equation \eqref{photonschrodinger}, and all of that is entirely relativistic.  The claims about probability density and probability current, like those in \eqref{photonprobdensity} and \eqref{photoncurrentdensity}, are not fundamental posits of the theory.  They should be derived from the dynamics of the wave function, along with some (hopefully true) assumptions about probability and/or rationality (see, e.g., \citealp{wallace2012, carrollsebens2014, barrett2017, sebenscarroll2018}).  If these derived probabilities don't transform properly, that seems problematic---even if the fundamental dynamics are perfectly relativistic.}

It is worthwhile to compare the status of the electromagnetic field to the complex scalar Klein-Gordon field.  Like Maxwell's equations, the Klein-Gordon equation can be interpreted as giving the dynamics for a classical relativistic field.  However, the Klein-Gordon field cannot be interpreted as a single-particle relativistic quantum wave function.  The Klein-Gordon field has a conserved four-current, but the density that features in this current is not always positive.  Thus, the density can be interpreted as a charge density, but cannot be interpreted as a probability density.\footnote{See, e.g., \citet[pg.\ 884--888]{messiah1962}; \citet[sec.\ 2-1-1]{itzyksonzuber1980}; \citet[sec.\ 3.1 and 3.4]{hatfield}; \citet[sec.\ 2.2]{ryder1996}.}  Both the electromagnetic and Klein-Gordon fields resist interpretation as relativistic wave functions because of problems with the probabilities derived from them.

Having reached the core lesson of this section, let us consider its bearing on the prospects for developing a Bohmian dynamics for the photon.  Clearly the probability assignments are problematically unrelativistic.  But, one might wonder if---despite this---the guidance equation acts relativistically.  If you use the guidance equation to calculate the photon velocity in one frame, there are two ways to find the velocity in another frame.  One can either Lorentz boost the state and reapply the guidance equation in the new frame (at the particle's location).  Or, one can simply boost the velocity.  However, these two methods will not always agree.\footnote{\citet[pg.\ 235]{bohmhiley} use a case in which the electric and magnetic fields are parallel and we boost along the axis they pick out to argue that the two above methods will not generally agree for the guidance equation derived from the Weber vector \eqref{badphotonguidance}.}  For the simple single plane wave state in \eqref{simpleplanewavefield} and \eqref{simpleplanewavestate}, using either of the two guidance equations proposed above, \eqref{photonguidance} or \eqref{badphotonguidance}, under either an $x$-boost or a $z$-boost, the two methods for finding the new velocity will yield the same result.  The Bohmian photon will move at $c$ in the direction of wave propagation.  For the superposition of two plane waves in \eqref{twowavestate} under a $z$-boost, the two methods of transforming the velocity will agree if one uses the guidance equation derived from the photon wave function \eqref{photonguidance}.  However, they will disagree if one uses the guidance equation derived from the Weber vector \eqref{badphotonguidance}.\footnote{According to \eqref{badphotonguidance},  a photon in \eqref{twowavestate} will not be moving at all (because there is no energy flow).  Thus, from the perspective of an observer moving with velocity $u\hat{z}$, it should be moving backwards with velocity $-u\hat{z}$.  However, if you apply \eqref{badphotonguidance} to \eqref{twowaveLT} you get a different photon velocity.}  Here the new guidance equation \eqref{photonguidance} works better than the old \eqref{badphotonguidance}.  However, for an $x$-boost of \eqref{twowavestate}---the case which quashed the hope that the probability density and current might together transform as a four-vector---two methods of transforming the velocity will disagree for either guidance equation.  So, we have not found a suitable relativistic guidance equation for the photon.

\section{Conclusion}

Dirac's wave equation for the electron can be interpreted either as giving the dynamics of a classical field or the dynamics of a quantum wave function.  Both the classical field theory and the quantum particle theory are relativistic.  In a similar manner, it is possible to interpret Maxwell's equations either as giving the dynamics of the classical electromagnetic field or as giving the dynamics of a quantum wave function for the photon.  However, in this case the classical field theory is relativistic but the quantum particle theory is not (the probabilisitic predictions for the results of position measurements are frame-dependent).  It is possible that there exists a relativistic quantum theory for the photon hidden within Maxwell's equations which we have not yet found.  Indeed, I think it is worthwhile to continue to search for such a theory---utilizing the examples in section \ref{EXsection} showing where a couple of the most straightforward attempts start running into trouble.\footnote{Alternatively, instead of looking within the equations of classical electromagnetism, one could attempt to find a relativistic quantum particle theory for the photon by proposing new equations (see, e.g., \citealp{kiessling2017}).  Then, one would need to either integrate these equations into a non-standard way of understanding our existing quantum field theory for the photon, or, put forward and defend an appropriately modified quantum field theory based on this new quantum theory of the photon.}  Still, the quantum theory based on Good's photon wave function is well-motivated and closely parallels Dirac's quantum theory of the electron (see table \ref{table1}).  As this theory is not relativistic, I suspect that Maxwell's equations cannot be interpreted as giving the dynamics for a relativistic quantum theory of the photon.\footnote{There is reasonably wide consensus that such an interpretation is unavailable, though the reasons given for this conclusion are varied.  See the references in \citet{kiessling2017}.}

In the introduction, we discussed the question of whether we should take a field or a particle approach to our quantum field theories of the electron and the photon.  Without a relativistic quantum particle theory for the photon, the particle approach appears to be unavailable.  So, we must take a field approach.  If you were then to defend a particle approach for the electron, you would be treating the electron and the photon very differently---as is standardly done in classical electromagnetism.  This strategy has been taken\footnote{\citet{bohmhiley} treat bosons as fields and fermions as particles in their chapters on extending Bohmian mechanics to relativistic quantum field theory.} and it has its appeals.  There are a number of reasons to prefer a particle approach over a field approach for the electron.  However, I think that it would be best if we could interpret quantum field theory in a unified way, as fundamentally either a theory of just particles or just fields.  I take the similarity between the mathematics for the photon and the electron (made manifest in section \ref{PWFsection})\footnote{From a field perspective, section \ref{PWFsection} shows that you can make electromagnetism look very similar to classical Dirac field theory by rewriting the equations of electromagnetism in terms of $\phi$ (which does not have to be interpreted as a quantum wave function; it can be seen instead as an unusual way of writing the state of the classical electromagnetic field).} to support this idea that we ought to adopt the same approach for both.  Because we cannot take a particle approach to the photon, this line of reasoning points towards taking a field approach for the electron---viewing the Dirac field as more fundamental than the electron.\footnote{Along similar lines, others have noted that the fact that we cannot interpret the Klein-Gordon field as a single-particle relativistic wave function (discussed in section \ref{EXsection}) suggests that we ought to interpret both the Klein-Gordon and Dirac equations as giving the dynamics for classical fields when we are using them to build quantum field theories.  See \citet[ch.\ 4]{ryder1996}; cf.\ \citet[sec.\ 3]{flemingbutterfield1999}.}  But, this is only one consideration in a larger debate.

\vspace*{12 pt}
\noindent
\textbf{Acknowledgments}
Thank you to Steve Carlip, Sheldon Goldstein, Chris Hitchcock, Michael Kiessling, Matthias Lienert, Ward Struyve, A. Shadi Tahvildar-Zadeh, Roderich Tumulka, David Wallace, and an anonymous referee for helpful feedback and discussion.  This project was supported in part by funding from the President's Research Fellowships in the Humanities, University of California (for research conducted while at the University of California, San Diego).

\end{document}